# A CMOS compatible microring-based on-chip isolator with 18dB optical isolation

**Li Fan, Jian Wang, Hao Shen, Leo T. Varghese, Ben Niu, Jing Ouyang, and Minghao Qi**
*School of Electrical and Computer Engineering and Birck Nanotechnology Center, Purdue University, West Lafayette, IN 47906*
lfan@purdue.edu, wang381@purdue.edu and mqi@purdue.edu

**Abstract:** We demonstrate strong optical nonreciprocity in microring add-drop filters with asymmetric input and output coupling coefficients. Up to 18dB isolation was achieved with a silicon-on-insulator high-Q microring of 5 micrometer radius.
© 2010 Optical Society of America
**OCIS codes**: (230.3240) Isolators; (230.5750) Resonators; (220.4241) Nanostructure fabrication.

## 1. Introduction

Optical isolation is an important functionality in optical information processing. However, compact optical isolators, similar to the diodes in electronic integrated circuits, has not been experimentally demonstrated in a CMOS compatible platform. Previously investigated monolithic isolators include magnetic garnet films [1] and single-sideband electro-optic modulator in III/V material [2], which are still large in size and not easy to incorporate in CMOS processing. Recently, new ideas were proposed to use time-varying media [3] or optomechanical effects [4]. Additionally, optical isolators were proposed to use nonlinearity in optical cavities [5]. However, such isolation assumes different input power for forward and backward propagation, which the proposed structure does not provide.

Here we show that the difference in power coupled to the optical cavity can be achieved with asymmetric coupling in microring based add-drop filters. Figure 1 shows the basic structure, in which the ring is placed closer to the straight waveguide than it is to the waveguide with a 180° bend.

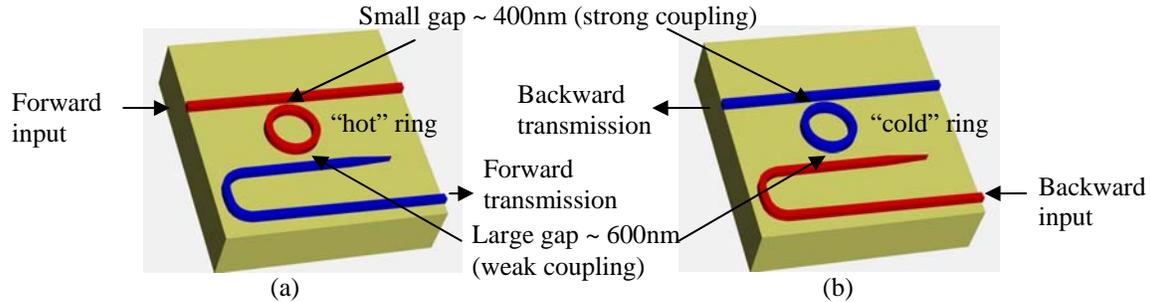

Fig.1: Schematic of the proposed ultra-compact optical isolator. (a) Strong forward input leads to high optical power in the input waveguide and in the ring, and red-shifts the resonance wavelength. (b) Strong backward input will not heat up the ring due to the low coupling coefficient.

If the waveguides and ring are made from strictly linear materials, the forward and backward transmission coefficient will be identical due to the reciprocity of the Maxwell's equations. However, silicon has strong thermo-optic coefficient and two-photon absorption (TPA). When a strong input is fed along the forward direction (Fig. 1a), more optical power will be coupled into the ring, and induces TPA. The generated free careers will absorb light and heat up the ring. Consequently, the resonance wavelength will be red shifted. On the other hand, when the same amount of optical power is launched backward into the ring, due to the large separation between the ring and the waveguide, less power will be coupled into the ring, and little heat will be generated thus little or no resonance wavelength shift.

## 2. Results and Discussion

We fabricated microring resonators on an SOI wafer with 250nm top-layer Si on 3 μm buried oxide. The radii of the rings are around 5 μm. The width of both the ring and the coupling waveguide is 500nm. The gaps between the waveguides and ring are set to be around 400nm for strong coupling and around 600nm for weak coupling. FDTD simulation shows roughly a 9× difference in coupling coefficients. The predominant optical mode is quasi-TM.

The device was first characterized in low input power (0.1mW fiber input and ~0.01mW when entering the ring coupling region) using an Agilent tunable laser source (TLS), and the forward and backward transmission was found to overlap almost completely, including the Fabry-Perot (FP) fringes. This verified the optical reciprocity for linear optical system. The loaded quality factor was determined to be 40,000.

For high input power levels, we first used the TLS in stepping mode, where the wavelength scanning can be arbitrarily slow, and laser transient power surge was excluded from the photodiode sampling window. This allows us to get consistent transmission spectra regardless of the scanning speed and direction, as well as the averaging time for the photodetector.

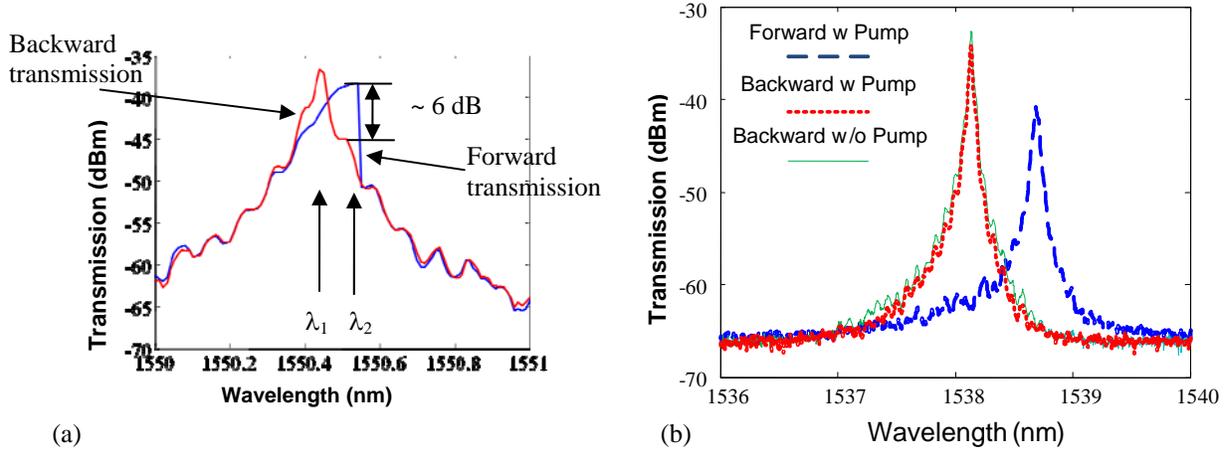

Fig.2: (a) Forward and backward transmission for an add-drop filter with coupling gaps of 400nm/600nm at about 1mW input power level at the ring. (b) Forward and backward transmission for an add-drop filter with appropriately detuned pump at $\lambda_2$ (~2mW optical power level at the ring).

Figure 2a shows the forward and backward transmission spectra when the input power is around 1mW (10mW fiber input) near the ring coupling region. A clear resonance peak shift around 0.1nm is observed. If one operates at $\lambda_2$ then an optical isolation of >6dB can be achieved between forward and backward transmission.

The achieved 6dB isolation, while reasonable for one ring, is not sufficient for device applications. Moreover, the isolation is input power dependent, which may further constrain its applicability. To enhance the isolation, and also to reduce the input power dependence, we can introduce a pump, located at $\lambda_2$ in Fig. 2a. In the case of forward transmission, according to Fig. 2a, $\lambda_2$ resonates with the ring. Therefore significant optical power exists in the ring and it becomes "hot". On the other hand, in the case of backward transmission, $\lambda_2$ is not in resonance with the ring, therefore low optical power exists in the ring and the ring remains "cold". We expect that there is about 30 fold difference in optical power stored in the ring when the pump is launched in different directions.

To further increase the resonance wavelength separation in forward and backward propagation, we increased the pump power at $\lambda_2$ to ~2 mW when it enters the ring. This will further heat up the ring and drive its resonance wavelength further into the long wavelength direction. Fig. 2b shows the transmission spectra near the resonance at a different FSR. Clearly, the forward transmission resonance (blue curve) is shifted about 0.6 nm to the right of the backward transmission (red curve). This separation is larger than the 50 GHz channel spacing that some of the DWDM scheme adopts. We also note that the backward transmission is not affected whether the pump is present.

Two operation modes can be envisioned: one is to transmit the data from the weak coupling port to the strong coupling port at 1538.2nm (backward direction in Fig. 1b), while block the opposite transmission. Fig. 2b shows that an isolation > 25dB (close to 30dB) can be achieved. The extra insertion loss is about 15dB. The other operation mode is to transmit the data from the strong coupling port to the weak coupling one at 1538.7nm (forward direction in Fig. 1a), while blocking the opposite direction. This would also provide an isolation close to 25dB, while the insertion loss will be slightly higher. However, the –3dB transmission of the blue peak is slightly larger, at about 0.064nm, thus in principle allows a data rate of 8GHz.

Figure 3 shows the independence of non-reciprocal transmission with regard to input power in the pump-assisted setup (pump power ~ 20mW). The different colors indicate the different fiber input power level. It is clear that the forward and backward transmission resonances are both independent of input power, which is across a 12dB dynamic range. We note that our scenario is different from the traditional pump-probe measurement, where the probe is much weaker than the pump. In our case, the input can be nearly as strong as the pump (10mW input vs 20mW pump at the fiber tip).

To further understand this finding, we conducted the bi-stability measurement in our system for the forward propagation. The input wavelength was slightly detuned to the long wavelength side by 0.05nm with regard to the "cold" resonance. Fig. 3b shows the traces of the output power when the input was increasing (red curve) and decreasing (blue curve). A clear bi-stability loop is shown. Moreover, when input power increased from 100mW to 150mW, the output power remained the same, and when it decreased from 100mW to around 25mW, the output power was also roughly constant. Intuitively, the increase in input power will cause the output power to increase. However, the higher power inside the ring will push the resonance further into the longer wavelength direction, and cause it to deviate from the input wavelength. In other words, the input wavelength will no longer be at the resonance wavelength, therefore causing the output to drop. These two effects will be balanced for certain input power levels and the output remains constant. This helps explain the independence of the transmission with regard to the input power in Fig. 3a, as the slight increase of input power does not change the output.

This effect might also allow one to design an optical limiter where the output could be equalized to a certain level for certain optical processing purposes, especially when nonlinearity is employed.

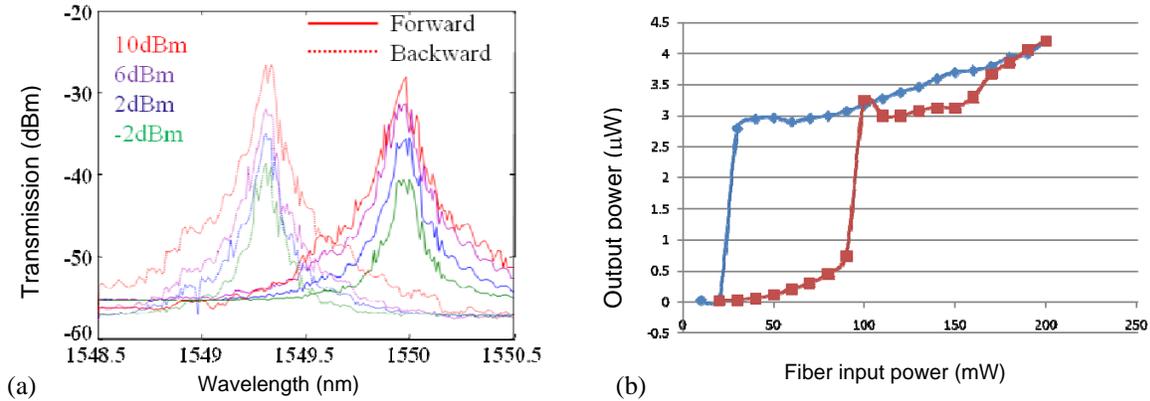

Fig.3: (a) Illustration of the input power independence for the pump-assisted non-reciprocal transmission. The solid lines are forward transmission spectra and the dashed lines are the backward transmission spectra. The red, violet, blue and green color correspond to the fiber input power levels of 10dBm, 6dBm, 2dBm, and -2dBm, respectively. The resonance wavelengths for the forward and backward transmissions are independent of input power. The pump power was fixed at 20mW for both forward and backward pumping. (b) Bi-stability curves for the asymmetric add-drop filter at the forward transmission condition. The red curve is the up-sweeping, where the input power increases, and the blue curve is the down-sweeping, where the input power decreases. There are two flat regions, 100mW-150mW for the up-sweeping and 25mW-100mW for the down-sweeping.

The bi-stability offers another intriguing application for isolation without pump: If the ring operates along the higher trace of the bi-stability loop, isolation could be improved in comparison to that shown in Fig. 1a. In a similar device, we first turned up the input power to 15mW near the coupling region in the forward direction, and recorded 8.1±0.4μW of output power at 1554.3nm. We then reduced the input power all the way down to 3mW near the coupling region, and the output was 6.8±0.4μW. Almost immediately after turning the input power below 3mW, the output quickly dropped down to around 40nW, which indicates a transition from the upper trace of the bi-stability loop to the lower one. After we switched the input/output, no such bi-stability was observed, and the output power was 0.11±0.2μW at 3mW input level. Therefore an isolation of 18±1dB was achieved without the assistance of a pump laser.

### 3. Conclusion

To our best knowledge, we have demonstrated the first CMOS compatible on-chip isolator, with an ultra-compact footprint of 15 μm × 15 μm. 18dB isolation was achieved with self-pumping and > 25dB isolation was achieved with a properly tuned pump. Preliminary optical limiting effect was also observed. We have also tested similar scheme in high-Q 1D photonic crystal cavity and achieved slightly higher isolation.